\documentclass[aps,pra,reprint,showkeys,nofootinbib]{revtex4-1}

\usepackage{pgf,pgfnodes,pgfarrows}
\usepackage{rotating}

\input{Qcircuit}

\usepackage{amsmath,amssymb,graphicx,dcolumn,bm,bbm,url}
\usepackage[colorlinks=true,linkcolor=red]{hyperref}
\newcommand{\arxiv}[2][arxiv:]{\href{http://arxiv.org/abs/#1#2}{#1#2}}
\newcommand{\arxivO}[2][quant-ph/]{\href{http://arxiv.org/abs/#1#2}{#1#2}}

\newcommand{\be}{\begin{equation}}
\newcommand{\ee}{\end{equation}}
\newcommand{\ben}{\begin{eqnarray}}
\newcommand{\een}{\end{eqnarray}}
\newcommand{\bes}{\begin{subequations}}
\newcommand{\ees}{\end{subequations}}
\newcommand{\bF}{\begin{figure}}
\newcommand{\eF}{\end{figure}}

\newcommand{\avg}[1]{\langle #1 \rangle}
\def\ket#1{ | #1 \rangle}

\def\tr{ {\rm{Tr }}\,}
\newcommand{\proj}[1]{\mbox{$|#1\rangle \!\langle #1 |$}}

\newtheorem{theorem}{Theorem}

\def\p{\textsf{P}}
\def\bqp{\textsf{BQP}}
\def\sp{\#\textsf{P}}
\def\np{\textsf{NP}}
\def\bpp{\textsf{BPP}}
\def\dqc{\textsf{DQC1}}
\def\nc{\textsf{NC1}}

\begin{document}

\title{Quantum Discord and Quantum Computing - An Appraisal}

\author{Animesh Datta}
\email{animesh.datta@physics.ox.ac.uk}
\affiliation{Clarendon Laboratory, Department of Physics, University of Oxford, OX1 3PU, United Kingdom}

\author{Anil Shaji}
\email{shaji@iisertvm.ac.in}
\affiliation{School of Physics, IISER TVM, Thiruvananthapuram, Kerala. India}

\begin{abstract}
We discuss models of computing that are beyond classical. The primary motivation is to unearth the cause of nonclassical advantages in computation. 
Completeness results from computational complexity theory lead to the identification of very disparate problems, and offer a kaleidoscopic view into the realm of quantum enhancements in computation. Emphasis is placed on the `power of one qubit' model, and the boundary between quantum and classical correlations as delineated by quantum discord. A recent result by Eastin on the role of this boundary in the efficient classical simulation of quantum computation is discussed. Perceived drawbacks in the interpretation of quantum discord as a relevant certificate of quantum enhancements are addressed.
\end{abstract}

\keywords{Mixed-state quantum computation, Complexity, Quantum discord, \dqc.}

\maketitle

\section{Introduction}

There is no known algorithm that may be implemented on a classical computer that can estimate the trace of a random $2^{n} \times 2^{n}$ unitary matrix efficiently. The scaling with $n$ of the number of steps taken by the best known classical algorithm to find the trace is exponential. However, there is an algorithm that works on a very restricted version of a quantum computer~\cite{kl98} that can estimate the normalized trace up to a fixed accuracy in a number of steps that is independent of $n$. The specialized quantum processor that can run the algorithm has precious little that may be identified as ``quantum'' in the traditional sense. Out of the $n+1$ qubits used, only one needs any amount of purity in its state. All the remaining qubits are in the completely mixed state. There is no entanglement between the qubit with purity and the rest at any step in the algorithm. The ``power of one qubit'' model, as this processor is known, is in the complexity class \dqc~\cite{s06} and is by no means a universal quantum computer. In this paper, the \dqc~model is of interest because it is one of the simplest models in which mixed quantum states lead to an exponential improvement over a classical computation. The aim here is to build a convincing case for considering non-classical correlations other than entanglement as an enabler of computational paths that a classical computer cannot take leading to a solution for some computational problems in an efficient manner.

Quantum entanglement is also as a kind of non-classical correlation; but not the only kind. It has provided deep insights into the role of non-classicality as a resource~\cite{b02,pv07,hhhh09} in quantum information processing. In particular, Josza and Linden~\cite{jl03} showed that the presence of multi-partite entanglement involving a number of parties increasing unboundedly with input size is necessary for a quantum algorithm {\em operating on pure quantum states} to offer an exponential speed-up over classical computation. If the entanglement does not increase unboundedly in a computational process, then it can be efficiently simulated classically using a classical computer to within any prescribed tolerance. An explicit construction for such a simulation scheme is also known~\cite{vidal03b}. On the other hand, quantum computations involving a restricted set of highly entangled states, such as stabilizer states, can be simulated efficiently classically, as shown by the Gottesman-Knill theorem~\cite{NC00}. Thus entanglement appears to be necessary but not sufficient for exponential speedups in quantum algorithms implemented using pure quantum states.

The fundamental problem that we endeavour to address in this paper is of identifying the resource or resources whose presence, individually or in groups, in the quantum states of a quantum information processor that enable it to perform tasks {\em exponentially} faster than the best known classical algorithm running on a classical information processing device. We will not be looking at sub-exponential speedups provided by quantum algorithms, as is the case with Grover's search algorithm~\cite{m00} and the Bernstein-Vazirani algorithm~\cite{bv93}, highly interesting as they are. The fundamental question raised above is by no means answered in this paper. Instead, the paper is meant to be a survey of what is known, and aimed to be an attempt to throw some light on how the problem may be approached. Of particular interest is the dismantling of the rather artificial division between quantum information processors that work using pure quantum states and those that work using mixed quantum states, and to identify potential candidates, unlike entanglement, that may be treated as uniform resources behind the exponential speed ups (if any) provided by both types.


\section{Not quite universal quantum computers}

While a result along the lines of Josza and Linden proving that a quantum resource is necessary and may be even sufficient for mixed state quantum computation is very much desirable, such a result has been quite elusive as well. To simplify the problem, we consider information processors that are not universal, yet perform particular computational tasks exponentially faster than the best known classical analogue.
These quantum computers do not usually satisfy all the requirements listed by DiVincenzo~\cite{divincenzo00} as necessary for assembling a scalable, universal quantum computer. They, however, might be considerably easier to implement than the full-scale, pure-state quantum computer discussed by Jozsa and Linden and hence they might see practical realization sooner. In the remainder of this section, we discuss a few examples of such limited quantum information processors which can work with mixed or pure quantum states to find solutions to specific problems efficiently.

Suppose we are allowed $n$ identical, non-interacting bosons to use in our computation. In particular, we have access to only single-photon input states, and photon-number-resolving measurements. This is certainly a subset of the requirements of DiVincenzo, but this model, called boson-sampling, is able to estimate the permanent of a $n \times n$ unitary matrix efficiently~\cite{aa10}. The permanent, like the determinant, is a polynomial in the entries of the matrix. Evaluating the permanent exactly is \sp-complete, a complexity class which can crudely be thought of as counting the number of solutions of a \np-complete problem~\cite{j03}. It is possible to estimate the permanent of a matrix with non-negative entries efficiently using Monte-Carlo Markov chains efficiently classically~\cite{jsv04}, but a unitary matrix is beyond the scope of these algorithms.

Another system of limited capabilities is a collection of quantum spins whose only mode of interacting is via their angular momentum~\cite{j10}. It has been shown to be able to approximate the irreducible representations of the symmetric group. There is no known classical algorithm for this problem, which is also related to the partition function of the Ponzano-Regge model, a 3-dimensional topological quantum field theory on a class of triangulated manifolds. This computational model, called permutational quantum computation, is particularly good for fault tolerance, as all operations are discrete. In fact, the set of $n+1$ spins interacting via their angular momenta has only $C_n$ coupling schemes, where $C_n$ is the $n^{\mathrm{th}}$ Catalan number. Thus, the number of possible two-spin interactions is $(n+1)!~C_n^2.$ Since
\begin{equation}
C_n = \frac{2n!}{n!(n+1)!},
\end{equation}
Stirling's approximation gives the total number of unitaries to be $\sim 2^{{\cal O}(n\log n)}.$  It is known that we need a number of unitaries doubly exponential in the number of particles involved for the universality of quantum computing. Thus, permutational quantum computation is likely to be less powerful than \bqp \footnote{BQP: Bounded error quantum polynomial - the class of problems that can be efficiently solved using a universal quantum computer. \p~is the class of problems that can be efficiently solved using a classical computer}, and is not known to include all of \p~either. The final limitation of this model, a practical one, is the inability to perform interferometric measurements to extract phase information. In spite of the limited power of the permutational model, it can approximate the partition function of a physical model for which there is no known efficient classical algorithm~\cite{j10}.

A third model of a specialized quantum information processor that is of particular interest to us is the  `power of one qubit' model. Here the normalized trace of a random unitary matrix is evaluated to any desired accuracy in a fixed number of steps independent of the size of the unitary. The computational process represented as a quantum circuit is given below:
\begin{equation*}
 \Qcircuit @C=1.5em @R=0.3 em {
  &  & \lstick{\ket{0}} & \gate{H} & \ctrl{1} & \qw & \measureD{X/Y} & \push{\rule{-1em}{4em}} \\
  &  & & \qw & \multigate{4}{U_n} & \qw & \qw \\
  &  & & \qw & \ghost{U_n} & \qw & \qw \\
  &  &\lstick{\mbox{$\frac{I_n}{2^n}$}} & \qw & \ghost{U_n} & \qw & \qw \\
  &  & & \qw & \ghost{U_n} & \qw & \qw \\
  &  &  & \qw & \ghost{U_n} & \qw & \qw \gategroup{2}{3}{6}{5}{0.5em}{\{}
}
\end{equation*}
The motivation for constructing this model comes from quantum computing using liquid-state nuclear magnetic resonance (NMR), where it is hard to prepare a number of pure qubits in a scalable manner, but easy to get a large numbers of them in a very mixed state.

The quantum state at the end of this circuit, just prior to the measurement, is given by
\ben
\rho_{n+1} &=& \frac{1}{2^{n+1}}\!\left[I_{n+1}+ \begin{pmatrix} 0&U_n\\ U_n^\dag&0 \end{pmatrix} \right] \nonumber\\
            &=& \frac{1}{2^{n+1}} \begin{pmatrix} I_n &   U_n \\ U_n^\dag & I_n \end{pmatrix},
\label{E:rhooutalpha}
\een
wherefrom it directly follows that $X$ and $Y$ measurements on the top qubit provide an estimate of the real and imaginary part of the normalized trace of the unitary, $\tr(U_n)/2^n.$ As the measurement is on a single qubit, each individual outcome is either $+1$ or $-1.$ An average over several trials provides the estimate. After $L$ tries, the estimate has an accuracy $\sim 1/\sqrt{L},$ given by the central limit theorem. The crucial point is that the error in the estimate is independent of the size of the problem, and this is what makes the estimation algorithm efficient. The algorithm can also be thought of as an averaged phase estimation algorithm. If the input to the lower $n$ qubits is the eigenvector of $U_n$ with eigenvalue $e^{i\phi_j},$  the outcome of $X$ and $Y$ measurements on the top qubit provides an estimate of $\cos\phi_j$ and $\sin\phi_j$ respectively. The completely mixed state is an equal mixture of all the eigenvectors, and the linearity of quantum mechanics immediately leads to the normalized trace, the sum of the eigenvalues.

The top qubit can be replaced by a mixed state given by,
\begin{equation}
\label{eq:sigma}
\sigma = \frac{1}{2}(I_2+\alpha Z).
\end{equation}
$X$ and $Y$ measurements lead to an estimate of $\alpha \tr(U_n)/2^n.$ This imposes an overhead of $1/\alpha^2$ for achieving the same precision in the estimate of the normalized trace as before. The overhead is tractable as long as $\alpha$ is not exponentially small in $n$. Therefore, we can get an exponential quantum speedup with the slightest bit of purity in the system. The model can thus be relabeled as the \emph{power of the tiniest fraction of a qubit}~\cite{datta05a}.

The power of one qubit model falls into the computational complexity class \dqc. The final state of the $n+1$ qubits in this model cannot be represented efficiently using the matrix product state formalism for mixed states\footnote{This algorithm~\cite{zv04} essentially vectorizes the density matrix, and looks for an efficient representation for that vector, given by a limited Schmidt rank.} in an efficient manner, thus excluding the possibility of classically efficient simulations of \dqc~computations using a large class of algorithms~\cite{datta07a}. The power of one qubit model has been exploited in a variety of problems, from estimating the density of states to the decay of fidelity in chaotic systems, with exponentially enhanced efficiency than the best known classical algorithms~\cite{pklo04,elpc04,eaohhk02,mpskln02,dlmp03,mrcl10}.

\section{The complexity class \dqc}

\begin{figure}
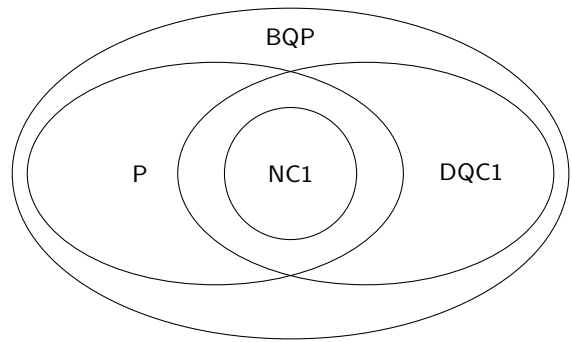

\centerline{
 \begin{pgfpicture}
        \pgfellipse[stroke]{\pgforigin}{\pgfxy(3.7,0)}{\pgfxy(0,2.2)}
        \pgfellipse[stroke]{\pgfpoint{1cm}{0cm}}{\pgfxy(2.5,0)}{\pgfxy(0,1.48)}
        \pgfellipse[stroke]{\pgfpoint{-1cm}{0cm}}{\pgfxy(2.5,0)}{\pgfxy(0,1.48)}
        \pgfcircle[stroke]{\pgfxy(0,0)}{25pt}
        \pgfputat{\pgfxy(0,0)}{\pgfbox[center,center]{\nc}}
        \pgfputat{\pgfxy(2.4,0)}{\pgfbox[center,center]{\dqc}}
        \pgfputat{\pgfxy(-2,0)}{\pgfbox[center,center]{\p}}
        \pgfputat{\pgfxy(0,1.8)}{\pgfbox[center,center]{\bqp}}
      \end{pgfpicture}}
\caption{The relative hierarchy of the complexity classes \p, \bqp~and \dqc. The \nc~class represents problems that can be solved efficiently using a classical parallel computer.}
\label{fig:complexityclass}
\end{figure}

The complexity class \dqc~represents the family of problems that may be solved efficiently with a restricted quantum information processor, the `power of one qubit' model.  This complexity class is known to be less powerful than \bqp~\cite{asv00}, and believed to be more powerful than \p. The relative inclusions of the different complexity classes is shown in Fig.~(\ref{fig:complexityclass}). There are several problems that have been identified to be in \dqc. We review a few of them before addressing the problem of the resources that make solving problems in this class using a non-universal quantum computer efficient.

The complexity of estimating the Jones polynomial of the plat closure of a braid at the fifth root of unity was shown to be BQP-complete~\cite{ajl06}. The same problem for the trace closure of a braid was shown to be \dqc-complete~\cite{sj08}, as was the estimation of its HOMFLY\footnote{The HOMFLY polynomial is a 2-variable generalization of the Jones polynomial~\cite{homfly}. The estimation of the HOMFLY polynomial of the plat closure of a braid was also shown to be \bqp-complete~\cite{wy06}.} polynomial~\cite{jw09}.  It has been known for quite some time that evaluating the partition functions of certain spin models, like the Ising model, is related to knot invariants, in particular, the Jones polynomial~\cite{w93}.  The difference between trace and plat closure is in the way the ends of the braids are joined to make a knot. This has a counterpart in the physical world. Trace closures correspond to periodic boundary conditions, while plat closures correspond to non-periodic boundary conditions~\cite{j89}. The partition function of a system depends on its boundary conditions. Therefore, one would expect that its estimation would behave analogously. For the Ising model defined on a lattice with periodic boundary conditions, estimating its partition function with polynomial accuracy was shown to be \dqc-hard~\cite{cueva11,al10}. Of course, estimation means approximation to its true value with a reasonable accuracy. The notion of `reasonable accuracy' is a fairly technical one~\cite{bflw05}. As the one-qubit model provides an efficient algorithm for an approximate solution to the estimation the Jones polynomial of the trace closure of a braid, it should be possible to approximate partition functions through the one-qubit algorithm and uncover physical reasons for the power of \dqc~model.

The complexity of computing matrix functions like traces, determinants, and permanents is an intriguing one. We have already seen that the estimation of the normalized trace of a unitary matrix is in \dqc. The permanent is believed to be extremely hard to evaluate, while the determinant of a matrix is easy to evaluate. Almost all the tractable counting problems are, in one way or another, reducible to the determinant of a polynomially sized matrix. In fact, it is always possible to write the permanent of one matrix as the determinant of another. The problem is that the size of the latter cannot be polynomially bounded in the size of the former. A similar dichotomy exists between the trace and plat closures of braids, the former of which is \dqc-complete, while the latter is \bqp-complete. As with matrices, it is always possible to transform the plat closure of a braid to the trace closure of another, but the error in the estimation depends exponentially on the number of strands in this case~\cite{ajl06}. This degrades the quality of the approximation of the trace closure exponentially.

Going beyond the physical and algebraic problems mentioned above, we now discuss a combinatorial problem, somewhat esoteric, which is in \dqc~and reveals connections to classical coding theory, graph theory, and through that, back to knot theory and statistical physics. The problem is that of evaluating quadratically signed weight enumerators (QSWEs)~\cite{kl01}. A general quadratically signed weight enumerator is of the form
 \begin{equation}
 \label{E:qswe}
S(A,B,x,y)= \sum_{b: Ab=0}(-1)^{b^T Bb}x^{|b|}y^{n-|b|},
  \end{equation}
where $A$ and $B$ are be matrices over $\mathbb{Z}_2$ with $B$ of dimension $n \times n$ and $A$ of dimension $m \times n$. The variable $b$ in the summand ranges over column vectors in $\mathbb{Z}_2$ of dimension $n$, $b^T$ denotes the transpose of $b$, $|b|$ is the weight of $b$ (the number of ones in the vector $b$), and all calculations involving $A$, $B$ and $b$ are modulo 2. The absolute value of $S(A,B, x, y)$ is bounded by $(|x| + |y|)^n$. In general, one can consider the computational problem of evaluating the sum in Eq.~(\ref{E:qswe}). In particular, it is known that for integers $k,l$, evaluating $S(A,B,k,l)$ exactly is $\sp$ complete. Now, let $A$ be square, of size $n \times n$, and let $\overleftarrow{A}$ denote the lower triangular part of $A$, which is the matrix obtained from $A$ by setting to zero all the entries on or above the diagonal. Let $diag(A)$ denote the diagonal matrix whose diagonal is the same as that of $A$ and $I$ denote the identity matrix. For matrices $C$ and $D$ with the same number of
columns, let $[C;D]$ denote the matrix obtained by placing $C$ above $D$. Then, the following two theorems hold~\cite{kl01}
\begin{theorem}
Given $diag(A)=I$, $k,l$ positive integers, and the promise that $|S(A,\overleftarrow{A},k,l)|\geq (k^2+l^2)^{n/2}/2$, determining the sign of
$S(A,\overleftarrow{A},k,l)$ is BQP-complete.
\end{theorem}

\begin{theorem}
Given $diag(A)=I$, $k,l$ positive integers, and the promise that $|S([A;A^T],\overleftarrow{A},k,l)|\geq (k^2+l^2)^{n/2}/2$, the sign of $S([A;A^T],\overleftarrow{A},k,l)$ can be efficiently determined by the the power of one qubit model.
 \end{theorem}
In these two problems, the integers $k$ and $l$ can be restricted to 4 and 3, respectively, without affecting their hardness with respect to polynomial reductions (using classical deterministic algorithms). Note that in the second one, the question of \dqc~completeness is still open.

\begin{figure*}[!htb]
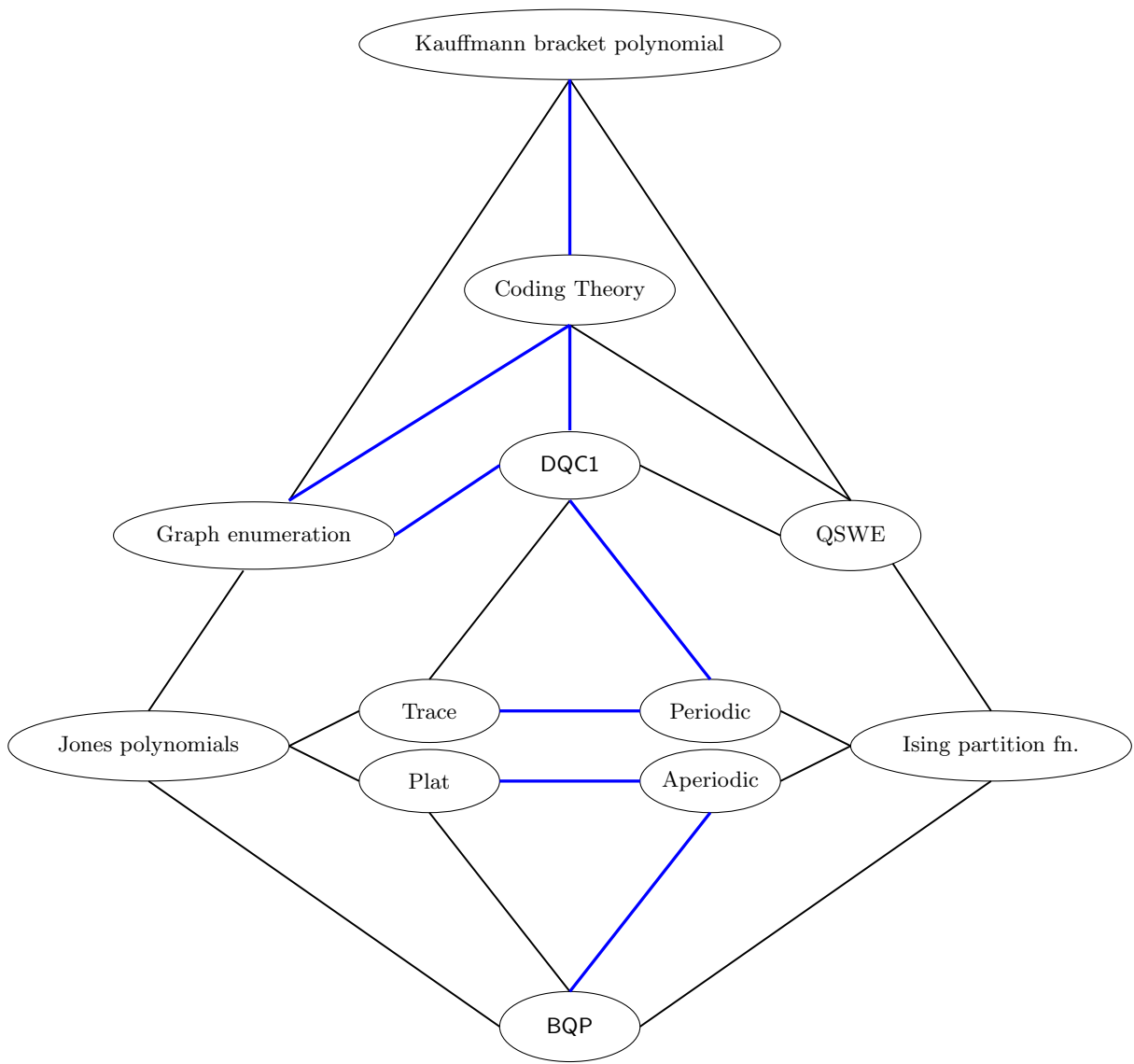

\centerline{
 \begin{pgfpicture}
        \pgfellipse[stroke]{\pgfpoint{0cm}{4cm}}{\pgfxy(3.0,0)}{\pgfxy(0,0.5)}
        \pgfellipse[stroke]{\pgfpoint{-6cm}{-6cm}}{\pgfxy(2,0)}{\pgfxy(0,0.5)}
        \pgfellipse[stroke]{\pgfpoint{0cm}{-2cm}}{\pgfxy(1,0)}{\pgfxy(0,0.48)}
        \pgfellipse[stroke]{\pgfpoint{4cm}{-3cm}}{\pgfxy(1,0)}{\pgfxy(0,0.50)}
        \pgfellipse[stroke]{\pgfpoint{6cm}{-6cm}}{\pgfxy(2,0)}{\pgfxy(0,0.5)}
        \pgfellipse[stroke]{\pgfpoint{-2cm}{-6.5cm}}{\pgfxy(1,0)}{\pgfxy(0,0.45)}
        \pgfellipse[stroke]{\pgfpoint{-2cm}{-5.5cm}}{\pgfxy(1,0)}{\pgfxy(0,0.45)}
        \pgfellipse[stroke]{\pgfpoint{2cm}{-6.5cm}}{\pgfxy(1,0)}{\pgfxy(0,0.45)}
        \pgfellipse[stroke]{\pgfpoint{2cm}{-5.5cm}}{\pgfxy(1,0)}{\pgfxy(0,0.45)}
        \pgfellipse[stroke]{\pgfpoint{0cm}{-10cm}}{\pgfxy(1,0)}{\pgfxy(0,0.5)}
        \pgfellipse[stroke]{\pgfpoint{0cm}{0.5cm}}{\pgfxy(1.5,0)}{\pgfxy(0,0.5)}
        \pgfellipse[stroke]{\pgfpoint{-4.5cm}{-3cm}}{\pgfxy(2,0)}{\pgfxy(0,0.48)}
    \pgfsetlinewidth{0.75pt}
        \pgfxyline(-4,-6)(-3,-6.5)          
        \pgfxyline(-4,-6)(-3,-5.5)          
        \pgfxyline(4,-6)(3,-6.5)            
        \pgfxyline(4,-6)(3,-5.5)            
        \pgfxyline(0,-9.5)(-2,-6.95)        
        \pgfxyline(0,-2.5)(-2,-5.05)        
        \pgfxyline(1,-2)(3,-3)              
        \pgfxyline(-1,-10)(-6,-6.5)         
        \pgfxyline(1,-10)(6,-6.5)           
        \pgfxyline(-4.65,-3.5)(-6,-5.5)      
        \pgfxyline(0,3.5)(-4,-2.5)          
        \pgfxyline(0,3.5)(4,-2.5)           
        \pgfxyline(4.6,-3.4)(6,-5.5)           
        \pgfxyline(0,0)(4,-2.5)             
        \pgfputat{\pgfxy(0,-2)}{\pgfbox[center,center]{\dqc}}
        \pgfputat{\pgfxy(4,-3)}{\pgfbox[center,center]{QSWE}}
        \pgfputat{\pgfxy(0,0.5)}{\pgfbox[center,center]{Coding Theory}}
        \pgfputat{\pgfxy(0,4)}{\pgfbox[center,center]{Kauffmann bracket polynomial}}
        \pgfputat{\pgfxy(-6,-6)}{\pgfbox[center,center]{Jones polynomials}}
        \pgfputat{\pgfxy(0,-10)}{\pgfbox[center,center]{\bqp}}
        \pgfputat{\pgfxy(-4.5,-3)}{\pgfbox[center,center]{Graph enumeration}}
        \pgfputat{\pgfxy(-2,-5.5)}{\pgfbox[center,center]{Trace}}
        \pgfputat{\pgfxy(-2,-6.5)}{\pgfbox[center,center]{Plat}}
        \pgfputat{\pgfxy(2,-5.5)}{\pgfbox[center,center]{Periodic}}
        \pgfputat{\pgfxy(2,-6.5)}{\pgfbox[center,center]{Aperiodic}}
        \pgfputat{\pgfxy(6,-6)}{\pgfbox[center,center]{Ising partition fn.}}
\pgfsetlinewidth{1.25pt}
    \color{blue}
        \pgfxyline(0,-9.5)(2,-6.95)          
        \pgfxyline(0,-2.5)(2,-5.05)          
        \pgfxyline(-1,-6.5)(1,-6.5)          
        \pgfxyline(-1,-5.5)(1,-5.5)          
\pgfsetlinewidth{1.25pt}
    \color{blue}
        \pgfxyline(0,3.5)(0,1)
        \pgfxyline(0,0)(0,-1.5)
        \pgfxyline(-2.5,-3)(-1,-2)          
        \pgfxyline(0,0)(-4,-2.5)            
      \end{pgfpicture}}
\caption{Problems that have connections to the \dqc~complexity class: Black lines show the connections already known in the literature. The blue lines show some of the possible connections that can be made. `Trace' and `Plat' refer to closures of braids whose Jones polynomials are of interest, while `Periodic' and `Aperiodic' refer to boundary conditions for Ising models. `Graph enumeration' refers to the class of graph theoretic problems like the evaluation of the Tutte, the dichromatic, or the chromatic polynomial. The Kauffman bracket polynomial is a knot invariant which is a generalization of the Jones polynomial, and is related to partition functions of more esoteric statistical mechanical models like the vertex model and the IRF (interactions round a face) model.}
\label{relations}
\end{figure*}

Lidar made the first connection between QSWEs and partition functions of spin models on certain classes of graphs~\cite{lidar04}. In that case, $A$ denotes the adjacency matrix of the relevant graph. Though these classes are highly restricted at present, attempts at broadening their scope are ongoing. Through the connection to QSWEs, \dqc~can be applied to analyze problems in coding theory, like weight generating functions of binary codes. Studying the complexity of such problems will certainly give us a better understanding of the complexity classes \p, \bqp, and \dqc~and their relative inclusions. As shown in Fig.~(\ref{relations}), the `power of one qubit' model can be employed to tackle problems in graph and knot theory, disciplines that have provided scores of problems for theoretical computer science in general and quantum computation in particular. Identifying problems in \dqc~and studying them in relation to \bqp-, \p-and \np-complete graph and knot theoretic problems might unravel the underlying mathematical structures that separate \p, \bqp, and \dqc. Such an understanding should illuminate greatly the mathematical framework underlying the power of quantum computation, just as studying the complexity of partition functions might unravel the physical foundations for that power. The diagram in Fig.~(\ref{relations}) outlines prospective avenues for exploring such problems in the context of mixed-state quantum computation, and has the potential of opening up completely new avenues in quantum computation research.

\section{Non-classical correlations in \dqc}

In the` power of one qubit' model discussed earlier, the top qubit is always separable from the rest~\cite{datta05a}. This lies at the centre of all the discourse involving the `power of the one qubit' model. If entanglement is the sole form of quantum correlations capable of providing quantum enhancements in computation, and there is none of it between the top qubit and the rest, how does the information travel from the bottom qubits to the top, allowing us estimate to the trace of the unitary that acts on the bottom qubits alone? Looking for other quantifiers of non-classical correlations in a quantum state is therefor a good starting point for addressing this problem.

Quantum discord was proposed as a measure of purely quantum correlations~\cite{z00,ollivier01a} about a decade ago. It was presented mathematically as a difference of two classically equivalent, but quantum mechanically inequivalent forms of the mutual information. The first is the quantum mutual information, which is a measure of all correlations - quantum as well as classical - in a quantum state. For two systems $A$ and $B,$ the mutual information is
\begin{equation}
\mathcal{I}(A:B) = H(A) + H(B)- H(A,B).
\end{equation}
If $A$ and $B$ are classical systems whose state is described by a probability distribution $p(A,B)$, then $H(\cdot)$ denotes the Shannon entropy, $H({\bm p})\equiv -\sum_j p_j \log p_j$, where $\bm p$ is a probability vector. If $A$ and $B$ are quantum systems described by a combined density matrix $\rho_{AB}$, then $H(\cdot)$ stands for the corresponding von Neumann entropy, $H (\rho) \equiv -\tr (\rho \log \rho)$.

For classical probability distributions, Bayes's rule leads to an equivalent expression for the mutual information, $\mathcal{I}(A:B) = H(B) - H(B|A)$, where the conditional entropy $H(B|A)$ is an average of Shannon entropies for $B$, conditioned on the alternatives for $A$.  For quantum systems, we can regard this form for $\mathcal{I}(A:B)$ as defining a conditional entropy, but it is not an average of von Neumann entropies and is not necessarily nonnegative. An alternative route to generalizing the classical conditional entropy to the quantum case is to recognize that classically $H(B|A)$ quantifies the ignorance about the system $B$ that remains if we make measurements to determine $A$.  When $A$ is a quantum system, the amount of information we can extract about it depends on the choice of measurement.  For a measurement given by a positive-operator-valued measurement (POVM) $\{\Pi_j\}$ corresponding to outcomes $j$, the state of $B$ after a measurement on $A$ is given by
\begin{equation}
\label{redstates}
 \rho_{B|j} = \tr_{\!A}\bigl(\Pi_j\rho_{AB}\Pi_j\bigr)/p_j, \;\; p_j=\tr_{A,B}\bigl(\rho_{AB}\Pi_j\bigr).
\end{equation}
A quantum analogue of the conditional entropy can then be defined as $\tilde{H}_{\{\Pi_j\}}(B|A) \equiv \sum_j p_j H(\rho_{B|j})\ge0$.  Since $\rho_B=\sum_jp_j\rho_{B|j}$, the concavity of von Neumann entropy implies that $H(B)\ge\tilde H_{\{\Pi_j\}}(B|A)$.  We can now define another quantum version of the mutual information,
\begin{equation}
 \label{quantJ}
\mathcal{J}_{\{\Pi_j\}}(A:B) \equiv H(B) - \tilde H_{\{\Pi_j\}} (B|A)\ge 0.
\end{equation}
Performing projective measurements onto a complete set of orthogonal states of $A$ effectively removes all nonclassical correlations between $B$ and $A$.  In the post-measurement state, mutually orthogonal states of $A$ are correlated with at most as many states of $B$. It is easy to see that these sorts of correlations can be present in an equivalent classical system. If on the other hand, the system $B$ is bigger than $A$, the
correlations are certainly non-classical, or quantum.

The value of $\mathcal{J}_{\{\Pi_j\}}(A:B)$ in Eq.~(\ref{quantJ}) depends on the choice of $\{\Pi_j\}$.  We want $\mathcal{J}_{\{\Pi_j\}}(A:B)$ to quantify {\em all\/} the classical correlations in $\rho_{AB}$, so we maximize $\mathcal{J}_{\{\Pi_j\}}(A:B)$ over all $\{\Pi_j\}$ and define a measurement-independent mutual information $\mathcal{J}(A:B) \equiv H(B) - \tilde H(B|A)\ge0$, where $\tilde H(B|A)\equiv \min_{\{\Pi_j\}}\sum_j p_j H(\rho_{B|j})$ is a measurement-independent conditional information. Henderson and Vedral~\cite{henderson01a} investigated how $\mathcal{J}(A:B)$
quantifies classical correlations, and an operational interpretation was also provided~\cite{dw04}.  The criteria they postulated for a measure of classical correlations $\mathcal{C}$ were
\begin{enumerate}
 \item $\mathcal{C}=0$ for product states
 \item $\mathcal{C}$ is invariant under local unitary transformations. This is because any change of basis should not affect the correlation between two subsystems.
 \item $\mathcal{C}$ is non-increasing under local operations. If the two subsystems evolve independently then the correlation between them cannot increase.
 \item $\mathcal{C}=H(\rho_A)=H(\rho_B)$ for pure states.
\end{enumerate}
One can then define a measure of purely quantum correlations as the difference of the total correlations in a system and $\mathcal{C}$. Taking the quantum mutual information to be a measure of total correlations in a system, the purely quantum correlations can be measured by
 \ben
\label{discorddef}
 \mathcal{D}(A,B) &\equiv& \mathcal{I}(A:B)-\mathcal{J}(A:B) \nonumber \\
                    &=&  \tilde H(B|A)-H(B|A).
\een
here is quite a substantial literature on the properties of quantum discord, its evaluation in a variety of systems, and its role on various phenomenon other than computation, and has been reviewed in this volume~\cite{revQD}. Quantum discord for pure states reduces to quantum entanglement, given by the von-Neumann entropy of the reduced state. Secondly, quantum discord is asymmetric between the two parties, $\mathcal{D}(A,B) \neq \mathcal{D}(B,A).$ Finally, the maximum value of quantum discord is the von-Neumann entropy of the measured subsystem, in this case $\mathcal{D}(A,B) \leq H(A).$ The set of states with zero discord will form an important ingredient in our discussion. There are several equivalent characterizations of this set, but an helpful one is that a state has zero discord if and only if it is block-diagonal in the marginal eigenbasis of the measured subsystem~\cite{facca10,datta10,hlz11,lcs11}.

It was proposed that quantum discord might be the resource behind the speedup offered by the power of one qubit model~\cite{datta08a}. The reason is two-fold. Firstly, there is no quantum entanglement between the top qubit and the rest, while there is a non-zero amount of quantum discord. Secondly, the amount of quantum discord across this split is a constant fraction of the maximum possible for this system, which is 1 ebit.  To better understand the role of discord in the \dqc~model, let us generalize the circuit slightly. The state of the top qubit is as in Eq.~(\ref{eq:sigma}) and it is not pure. The Hadamard gate $H$ is replaced by a general 1-qubit unitary $V$ given by
\begin{equation}
V = \left(
      \begin{array}{cc}
        e^{i\varphi}\cos\theta & e^{i\chi}\sin\theta \\
        -e^{i\chi}\sin\theta &  e^{-i\varphi}\cos\theta \\
      \end{array}
    \right).
\end{equation}
Then the input state is
$$
\frac{1}{2^{n+1}}\left(
             \begin{array}{cc}
               1+\alpha\cos2\theta & -\alpha e^{i\delta}\sin2\theta  \\
               -\alpha e^{-i\delta}\sin2\theta & 1-\alpha\cos2\theta \\
             \end{array}
           \right)\otimes I_n,
$$
where $\delta = \varphi + \chi,$ and the output state, in block form, is
\begin{equation}
\rho_{n+1}(\alpha) = \frac{1}{2^{n+1}}\left(
             \begin{array}{cc}
               (1+\alpha\cos2\theta)I_n & -(\alpha e^{i\delta}\sin2\theta)U_n  \\
               -(\alpha e^{-i\delta}\sin2\theta)U_n^{\dagger} & (1-\alpha\cos2\theta)I_n \\
             \end{array}
           \right).
\end{equation}
The circuit looks like
\begin{equation*}
\label{E:circuitalpha}
 \Qcircuit @C=1em @R=0.5em {
  & A & & & \lstick{\sigma} & \gate{V} & \ctrl{1} & \qw & \measureD{X/Y} & \push{\rule{-1em}{4em}} \\
  & B  & & & & \qw & \multigate{4}{U_n} & \qw & \qw \\
  &   & & & & \qw & \ghost{U_n} & \qw & \qw \\
  &  & & &\lstick{\mbox{$\frac{I_n}{2^n}$}} & \qw & \ghost{U_n} & \qw & \qw \\
  &   & & & & \qw & \ghost{U_n} & \qw & \qw \\
  &   & & &  & \qw & \ghost{U_n} & \qw & \qw \gategroup{2}{5}{6}{5}{0.5em}{\{} \gategroup{1}{2}{1}{9}{2em}{--} \gategroup{2}{2}{6}{9}{2em}{--}
}
\end{equation*}
and the state is completely separable across the split given by the dotted lines. If the unitary has the eigendecomposition $U_n = \sum_j e^{i\phi_j}\proj{e_j},$ the state can be written in the form
\begin{equation}
\rho_{n+1}(\alpha) = \frac{1}{2^{n+1}}\sum_j (\proj{\xi_j} + \proj{\zeta_j})\otimes \proj{e_j},
\end{equation}
where
$$
\ket{\xi_j} = a\ket{0} + e^{i(\phi_j+\delta)}b\ket{1},~~~\ket{\zeta_j} = c\ket{0} + e^{i(\phi_j+\delta)}d\ket{1},
$$
with the real parameters $a,b,c,d$ given by
\begin{widetext}
$$
a^2 + c^2 = 1+\alpha\cos2\theta,~~b^2 + d^2 = 1-\alpha\cos2\theta,~~ab + cd = \alpha\sin2\theta,~~a^2 + b^2 = 1 = c^2+d^2.
$$
\end{widetext}
The state is thus separable across the dotted boxes above.

\begin{figure}[!htb]
\begin{center}
\resizebox{8.5cm}{6.5cm}{\includegraphics{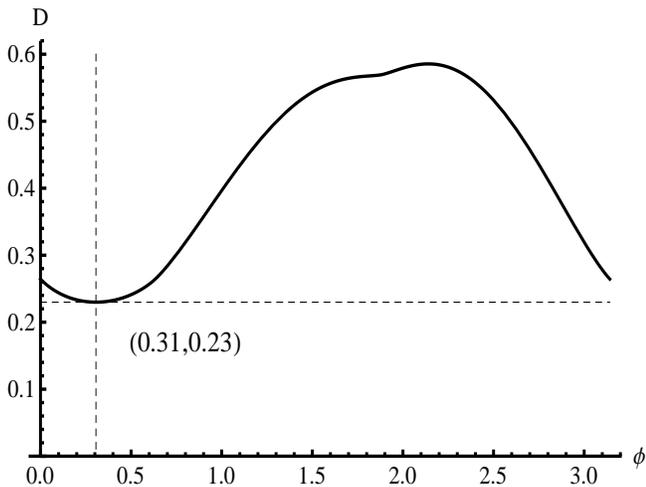}}
\caption{The quantum discord in the circuit that estimates the Jones polynomial of the trace closure 4-strand, with the first and the second strands crossing. The discord is $\mathcal{D} = 0.23,$ and we plot the difference of the mutual and conditional information as a function of the measurement
$\mathfrak{M}_A = \cos\phi X + \sin\phi Y.$ The optimal measurement is given by $\phi = 0.31.$}
\label{fig:knotdiscord}
\end{center}
\end{figure}

However, there is quantum discord across that split for a typical unitary~\cite{datta08a}, when the measurement is made on the subsystem $A$. One can show that the optimal measurement lies in the $X-Y$ plane, given by $\mathfrak{M}_A = \cos\phi X + \sin\phi Y$.  In  Fig.~(\ref{fig:knotdiscord}), the difference ${\cal I} - {\cal J}$ is plotted as a function of the measurement angle $\phi,$ for the unitary corresponding to the trace closure of a 4-strand braid~\cite{pmtl11}. The minimum of the graph, which gives us the value of discord is also shown.

The modified version of the power of one qubit model that we have considered here  provides approximations to the value the Jones polynomial of a simple knot at the fifth root of unity. A recent experiment\footnote{An all-optical experimental implementation of the \dqc~model has also been carried out~\cite{lbaw08}, as has been liquid-state NMR implementations to evaluate Jones polynomials~\cite{pmrl09,marx10}.} has confirmed that there is indeed non-zero discord in the \dqc~circuit that estimates the Jones polynomial of the trace closure of the 4-strand braid, with the first and the second strands crossing~\cite{pmtl11,dvb10}. In this experiment, the polarization of the top qubit was quite small, $\alpha \sim 10^{-5},$~\cite{gina11} which leads to a very small, but strictly non-zero value of discord. The plot in Fig.~(\ref{fig:knotdiscord}) is for $\alpha = 1$. The presence of discord in this circuit together with the absence of entanglement has led to the proposal of quantum discord being the resource for the enhanced power of the class \dqc. However, this is not a formal proof that discord is necessary for quantum enhancements in the \dqc~model.

As we mentioned earlier in this section, discord is asymmetric across the subsystems, and the quantum discord in $\rho_{n+1}(\alpha)$ for measurements made on subsystem $B$, given by the projectors onto the eigenvectors of $U_n,$ is zero. This can be used to repudiate the above claim of discord being a resource for the power of \dqc. To explore this further, let us perform a simple calculation. Say we measure the $n$ qubits in $B,$ using an operator $\mathfrak{M}_B.$ The expectation value of such a measurement is
\ben
\avg{\mathfrak{M}_B} &=& \tr\left(\rho_{n+1}(\alpha)\left(
                                            \begin{array}{cc}
                                              \mathfrak{M}_B & 0 \\
                                              0 & \mathfrak{M}_B \\
                                            \end{array}
                                          \right)\right) \nonumber \\ 
                    & = & \frac{\tr(\mathfrak{M}_B)}{2^n}.
\een
The result is independent of the input state $\sigma$ and the unitary $V$. This shows that no information about the system $A$ is obtainable by measuring system $B,$ no matter what measurement we make. Since any computation can be thought of as the flow of information, zero discord in one direction in the one qubit model merely reflects the just derived fact that there is no information propagating across the dotted boxes from $A$ to $B.$ Thus, what might be considered as a drawback for quantum discord, its asymmetry, can heuristically be understood as designating the direction of propagation of quantum information. Indeed, pure-state quantum computations are completely unitary, thereby making information transfer between parts of such a system necessarily symmetric. For such systems, quantum discord is identical to quantum entanglement which, as a measure of correlations, is also symmetric between the systems involved. Not so for a mixed-state quantum computer.  As in the power of one qubit model, states with non-zero quantum discord might provide a possible explanation to quantum computation with mixed quantum states, and the asymmetry of the measure could be vital to its veracity.

Recently, it has also been shown that almost all states possess non-zero quantum discord~\cite{facca10}. This is unlike entanglement, where the measure of the set of separable states is non-zero. In particular, there is a ball around the completely mixed state inside which all states are separable~\cite{gb03,gb05}. This has generated inordinate amounts of excitement, in its manifestation as the sudden death of entanglement. There is no such phenomenon for quantum discord, and the set of states with zero quantum discord is of measure zero. Consequently, one might wonder how so ubiquitous a quantity as quantum discord might be a resource.

One may build some intuition on the role of nonclassical correlations in quantum computing using the following heuristic: Consider a generic quantum information processor that uses registers of qubits for input and output, and  an additional register of qubits for the computation. Both the input and output are human-readable bit strings; the input bit string being loaded onto the qubit register typically by a process or quantum state preparation and the  quantum state of the main register at the output being reduced to a bit string by a measurement.  The initial state of the computing qubits may be chosen so that the starting state of the quantum computer has no nonclassical correlation, including discord, across any bipartite or multipartite division. The state at the output end may also be chosen likewise by measuring them if needed.

One can now imagine the computational process to be a sequence of discrete steps that takes the all the qubits from the initial, zero-discord, state to the final, zero-discord, state. The discrete steps correspond to the action of each gate set in the computational sequence. Since a comparison with classical computing is attempted wherein the computational steps are discrete transformations and not unitary gate actions, let us ignore what happens while each gate is acting. In other words, we consider only the states before and after the action of each gate. Imagine that we now run a classical algorithm as well as a quantum algorithm for the solving the same computational problem on the quantum information processor. The classical algorithm will demand that at each step the state of the quantum computer have no nonclassical correlation across any bipartite division. Note that a deterministic classical algorithm will be even more restrictive, demanding product states in the computational basis at each intermediate step. So we are including probabilistic classical algorithms as well. This means that one has to go from the initial to the final state jumping from one zero discord state to another without ever leaving this set of measure zero. A true quantum computation however, lets one take alternate paths involving intermediate states with nonclassical correlations which might result in reaching the desired output state in significantly - perhaps even exponentially - fewer computational steps.

\begin{figure}[!htb]
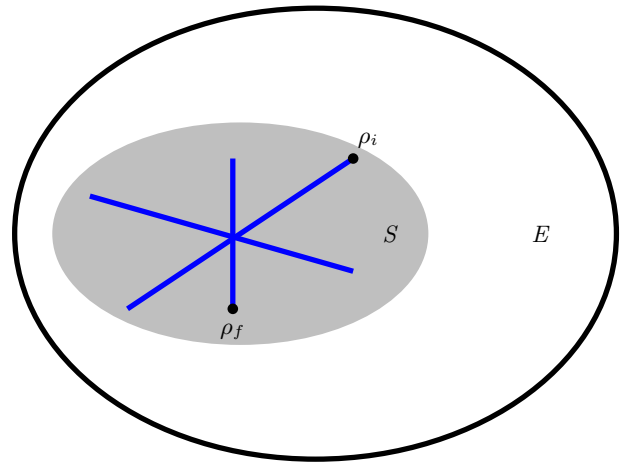

\centerline{
 \begin{pgfpicture}
        \pgfsetlinewidth{2pt}
        \pgfellipse[stroke]{\pgforigin}{\pgfxy(4,0)}{\pgfxy(0,3)}
        \color{lightgray}
        \pgfellipse[fill]{\pgfpoint{-1cm}{0cm}}{\pgfxy(2.5,0)}{\pgfxy(0,1.48)}
        \color{blue}
        \pgfxyline(-3,0.5)(0.5,-0.5)
        \pgfxyline(-2.5,-1)(0.5,1)
        \pgfxyline(-1.1,-1)(-1.1,1)
        \color{black}
        \pgfcircle[fill]{\pgfxy(0.5,1)}{2pt}
        \pgfcircle[fill]{\pgfxy(-1.1,-1)}{2pt}
        \pgfputat{\pgfxy(0.7,1.25)}{\pgfbox[center,center]{$\rho_i$}}
        \pgfputat{\pgfxy(-1.1,-1.3)}{\pgfbox[center,center]{$\rho_f$}}
        \pgfputat{\pgfxy(1,0)}{\pgfbox[center,center]{\textit{S}}}
        \pgfputat{\pgfxy(3,0)}{\pgfbox[center,center]{\textit{E}}}
        \end{pgfpicture}}
\caption{A schematic representation of the space of quantum states. The gray space $\textit{S}$ is the set of separable states, the rest is the set of entangled states $\textit{E}$. The larger ellipse is the set of all states. The blue lines are the set of zero discord states, and they intersect at the completely mixed state. If one is proceed from $\rho_i$ to $\rho_f$ touching down only on the blue lines, the computation might take more steps than the case where the state can explore the entire space.}
\label{fig:grid}
\end{figure}

This above heuristic is represented in Fig.~(\ref{fig:grid}). A classical algorithm will restrict the state of the quantum computer on the blue lines representing zero discord states. A quantum algorithm will not. The states $\rho_{i}$ and $\rho_{f}$ are the initial and target states of the quantum computer.  Attempts are ongoing to formalize this simple intuition. It is to be noted that this is more of a qualitative than a quantitative argument, and what appears to be important here is the distribution of zero discord, classical states in the space of all quantum states.

Eastin~\cite{e10} has recently proved a significant result along the lines described above. He considered a  quantum computation for which the input state is a product state diagonal in the computational basis, $\rho^0 =\otimes_k \rho_k^0$. The computation is done by a sequence of unitary gates $\{G^t\}$, and concluded by single-subsystem measurements determining the outcome of the computation. Eastin showed that if the quantum state of the qubits involved at every step of the computation is {\em concordant} then the computation can be simulated efficiently using a classical computer. A concordant state is one that is diagonal in a product basis.  This is equivalent~\cite{lcs11} to $\mathcal{D}(A,\bar{A}) = 0 = \mathcal{D}(\bar{A},A)$ for any subsystem $A$ and its complement $\bar{A}.$ In other words, the quantum discord of the state across any bipartition must be zero.

Eastin's result is restricted in the sense that it does not allow {\em any} nonclassical correlations, even those limited to a subset of qubits, during the computation. This is unlike the Joza-Linden result which showed that the computational steps can be efficiently simulated even if entanglement, restricted to a subset of qubits, is present. Eastin's result is also a qualitative one in which the presence and absence of discord is considered and not the amount of discord. In addition to this, Eastin defines quantum discord as a minimization over a set of complete, orthogonal set of one dimensional projectors~\cite{ollivier01a}, and not general POVMS~\cite{dattathesis}. The gate operations in the computation are restricted to one and two qubit gates as well. This restriction, if relaxed, could lead the way to a more quantitative version of Eastin's result, in that one could study the classical simulation of systems which are approximately concordant. It would also allow for studying the classical simulation of systems with quantum discord restricted to blocks of bounded size. This would be akin to the results of Vidal~\cite{vidal03b}, as extended to mixed states.

Finally, for completeness, we consider an argument that was raised against considering discord as a resource behind exponential speedups. Consider the \dqc~circuit when
\begin{equation}
\label{eq:zerodis}
U_n = k U_n^{\dagger}.
\end{equation}
This is a necessary and sufficient condition for vanishing discord in the pre-measurement state in the \dqc~model. Let us first note that $k= e^{i\vartheta},$ must be a phase. It has been suggested that for a unitary of this form, the \dqc~circuit provides an exponential speedup in the absence of quantum discord, and so quantum discord cannot be the source for the quantum advantages in this model~\cite{dvb10}. However, if we decompose the unitary $U_n$ in terms of any universal gate set, the above condition does not impose any restriction on the intermediate states of the computation. Therefore, quantum discord may very well be present in any of the states preceding the final one. As a simple example, consider $U_n = Z,$ the Pauli operator. This operator can be composed of two phase gates $P$, since $Z = P^2,$ where
\begin{equation}
P = \left(
        \begin{array}{cc}
          1 & 0 \\
          0 & i \\
        \end{array}
      \right).
\end{equation}
Although, the final \dqc~state has no quantum discord, the intermediate state after the application of one phase gate has non-zero quantum discord. Thus, zero quantum discord in the final state does not imply the absence of discord at all stages of the computation. Furthermore, zero quantum discord in every intermediate state does not imply the same at the final state. This is because the unitaries satisfying Eq.~(\ref{eq:zerodis}) do not form a group, implying that if we imposed this condition at each step of the decomposition, the final unitary need not, in general, have this property. In other words, if $U_n = W_1W_2\cdots W_T$, and each $W_i$ satisfies Eq.~(\ref{eq:zerodis}) (upto the phase factor $k$),
\begin{widetext}
\begin{equation}
U_n = W_1W_2\cdots W_T = W_1^{\dag}W_2^{\dag}\cdots W_T^{\dag} \neq (W_1W_2\cdots W_T)^{\dag} = U_n^{\dag}.
\end{equation}
\end{widetext}
It can thus be concluded that a \dqc-circuit with a unitary satisfying Eq. (\ref{eq:zerodis}) is not a valid rebuttal of the suggestion that quantum discord might be responsible for the speedup in the power of one qubit model.

\section{Conclusions}

Identifying the boundary between quantum states with nonclassical correlations and those without would significantly aid in designing information processing algorithms and devices that can provide exponential speedups over their classical analogues. We have reviewed the role of quantum discord as a quantifier of nonclassical correlations. We have presented some evidence that discord might be necessary for quantum speedups in mixed-state quantum computations. In particular, we noted that the formal proofs in this direction are more qualitative in nature than quantitative, and a direct connection between the amount of quantum discord in a system and its computational power is lacking. Parallel to developing the connection between quantum discord and computation, there have been advances in understanding the nature of quantum discord as well. Quantum discord has an operational interpretation in quantum information theory in terms of quantum state merging~\cite{md11,cabmpw11}, as well as in terms of the thermodynamics of quantum and classical Maxwell's demons~\cite{zurek03b,bt10}.

There are results~\cite{j10} that suggest that quantum discord is not sufficient for quantum speedups using mixed states. The \dqc~version of permutational quantum computation provides an estimate of the normalized character of the irreducible representations of the symmetric group. This problem is known to be in \bpp, since a randomized sampling algorithm can provide the answer, yet there is quantum discord in the circuit that implements it. Additionally, this is indirect evidence that permutational quantum computation is weaker than \dqc~\cite{j10}. There are no results yet on the relative strengths amongst classes of non-universal quantum computational models, and classes of problems like counting, decision, or sampling. Such results will be extremely valuable in understanding features of quantum mechanics, and their implications as quantum resources.

The unlikelihood of a single quantity certifying all forms of quantum advantages is becoming apparent with our increased understanding of quantum information science. Yet, we can attempt at formalizing the notion of quantumness into as few an independent number of quantities as possible for a given system or scenario. In course of this exercise, we must make every attempt to keep our system or scenario as generic as possible. What is progressively becoming clear is the presence of a trade-off between the conciseness of our understanding, and the scope of its validity. In the long run, the validity will be governed by the practical feasibility of the model. Given that the construction of a full-scale quantum computer is still a challenging prospect, intermediate models of computation are more than theoretical curiosities. If such models are able to provide (even polynomial) speedups over best known classical algorithms for certain tasks, it would be a significant advance in the field. More so if the tasks are widely relevant, for instance, matrix inversions, root-finding, Fourier transforms. Therefore, intermediate models of computation are a vital component in the quest for tangible quantum advantages. In the process, they will clarify what differentiates quantum from classical when it comes to computational advantages.

Quantum computing with mixed states seem to be a general enough setting for which we ought to find the certificates of quantum enhancement. This is more than a mere theoretical exercise with abstract implications. Substantial experimental effort in quantum computation is directed at maintaining the purity of the quantum systems. A better understanding of mixed-state quantum computations that provide exponential enhancements will help direct valuable resources in harnessing the available potential, rather than shying away from them merely because they involve states that are not pure. \dqc~is such a model. The promise of the \dqc~model lies in its strong interconnections to several areas of physics and mathematics. The set of complete problems for this class is varied, from knot theory to coding theory, and statistical mechanics to topological field theory~\cite{ajkw10,ja11}. This makes it an ideal workbench in the analysis and understanding of the power of different physical models and the computations of different invariants related to these models, as well as potential resources for quantum enhancements such as quantum discord. The interplay between the computational efforts needed to estimate elements versus characters of irreducible representations, individual matrix elements versus traces of unitary matrices, knot invariants of plat versus trace closures of braids, and partition functions of systems with periodic versus non-periodic boundary conditions on square lattices provide us with a variety of possibilities to unlock the true sources of quantum enhancements in computation, as indicated in Fig.~(\ref{relations}). These are all problems primed to be explored.

\section*{Acknowledgements}

This work has been informed by numerous discussions with several people including C.~M.~Caves,  S.~T.~Flammia, S. Boixo,  V. Madhok, W. H. Zurek, R. Jozsa, G. Adesso, A. Acin, C. Brukner, B. Eastin, K. Modi, A. Brodutch, C. A. Rodr\'iguez-Rosario, G. Passante, R. Laflamme, I. A. Walmsley. AD was funded in part by EPSRC (Grant EP/H03031X/1), the European Commission (FP7 Integrated Project Q-ESSENCE, grant 248095), and the US European Office of Aerospace Research and Development (Grant 093020). AS acknowledges the support of the Department of Science and Technology, Government of India through the Ramanujan Fellowship and the Fast Track Scheme for Young Scientists grant No:~00/IFD/5771/2010-11.

\end{document}